\documentclass[12pt]{article}
\usepackage{amstex}
\newcommand{\lsim}{\raisebox{-0.13cm}{~\shortstack{$<$ \\[-0.07cm] $\sim$}}~}

\newcommand{\defprod}{\raisebox{-0.13cm}{~\shortstack{$\prod$ \\[-0.07cm] 
${}_{j\neq i}$}}~}

\newcommand{\nn}{\noindent}

\def\beq{\begin{equation}}
\def\eeq{\end{equation}}

\def\beqn{\begin{eqnarray}}
\def\eeqn{\end{eqnarray}}
\relax
\def\ba{\begin{array}}
\def\ea{\end{array}}

\renewcommand{\theequation}{\thesection.\arabic{equation}}

\begin{document}

\begin{flushright}
PM/99--09 \\
\end{flushright}

\vspace{1cm}

\begin{center}

{\large\sc {\bf 
A convergent scheme for one-loop evolutions of the Yukawa couplings in the MSSM \\}} 
\vspace{1cm}
{ G. Auberson and G. Moultaka}
\vspace{1cm}

Physique Math\'ematique et Th\'eorique, UMR No 5825--CNRS, \\
Universit\'e Montpellier II, F--34095 Montpellier Cedex 5, France.

\end{center}

\vspace*{2cm}

\begin{abstract}

\noindent

Integrated forms of the one-loop evolution equations are given for the 
Yukawa couplings in the MSSM, valid for any value of $\tan \beta$,
generalizable to virtually any number of Yukawa fermions, and  
including all gauge couplings.  
These forms 
turn out to have nice mathematical convergence properties which we prove,
and we determine the ensuing convergence criteria. Furthermore,  they
allow to write down general sufficient and necessary conditions to 
avoid singularities in the evolution of the Yukawa couplings over physically 
relevant energy ranges. 
We also comment briefly  on the possible use of these features for physics 
issues and give a short numerical illustration.

\end{abstract}
\newpage

\section{Introduction}

Large Yukawa couplings often play an important role in spontaneous 
electroweak symmetry breaking models such as the standard model,
its extensions, and also in some alternatives to it.
The existence of Infrared (IR) attractive fixed points \cite{PR} and 
effective IR 
fixed points \cite{Hill} in such a regime can be a benchmark in selecting
the phenomenologically viable theories in view of the measured mass of
the top quark. Such considerations clearly favoured the minimal supersymmetric
extension of the standard model (MSSM) \cite{EWB}, with its upper bound on
the top mass, $m_t \leq (190 -200 GeV) \sin \beta$ \cite{Alvarez}
over alternatives such as \cite{BHL}. 

In supersymmetry there is of course much more to it.   
More specifically, in the minimal
supersymmetric extension of the standard model (MSSM) \cite{Review}, 
the spontaneous
breaking of the electroweak symmetry is generically driven by the large top 
quark mass in association with a soft supersymmetry breaking sector \cite{EWB}.
Such a possibility can be looked at as the phenomenological facet 
of a [still to be discovered] deep connection between the origin of 
supersymmetry breaking and that of the electroweak symmetry breaking. 
Meanwhile, it helps in correlating theoretically the very many free parameters 
of the MSSM, leading to quantitative estimates of the supersymmetric partners 
spectrum, of prime importance in guiding the experimental search for 
supersymmetry, and relating (at least qualitatively) the physics over as many 
orders of magnitudes as between the GUT scales and the electroweak scale.  

Of course, the key point in all the above issues is the way the various 
parameters ``run'', as dictated by the renormalization group equations. 
The RGE's are, in general, complicated coupled differential equations
already at the one-loop level and one usually resorts to numerical methods
to solve them \cite{spectrum}. However, analytical solutions would be desirable for
several reasons [besides the obvious one of allowing a better control over
the  structure of the running.] The radiative breaking of $SU_c(3)\times SU_L(2)
\times U_Y(1)$ and the structure of the vacuum are controlled  by the effective
potential (EP) of the theory. The theoretical improvement of the functional
form of the EP needs the analytical form of the running of involved
quantities such as masses, couplings and fields which enter the game. If the
analytical form of the EP beyond the tree-level were known exactly, or at least
in an RG improved form beyond the naive loop corrections, one could hope to
determine the relations the initial values of the parameters should fulfill,
to break the electro-weak symmetry at the right energy scale and avoid
in the same time color and charge breaking vacua\footnote{in the absence of 
such a knowledge, due partly (in the RG improvement program) to the existence
of many different mass scales, one relies on rough approximations hoping that
they encompass the leading behaviour \cite{casas}.}. Another important application 
of the analytical solutions is the determination of the effective behaviour
in the low energy regime (infra-red regions) independently (but within a
domain) of the actual values in the ultra-violet, such as the the top Yukawa
coupling effective fixed point \cite{Hill}, or the triviality bounds on the 
Higgs mass.

Analytical solutions of the complete set of RGE's in the MSSM were
known to one-loop order for small $\tan \beta$,
 strictly speaking in the case all Yukawa couplings are put to zero except for
the top quark. Actually the structure of the coupled equations is such that
a necessary condition to solve them entirely is to be able to solve first for
the Yukawa couplings. This is of course not enough, and initially one assumed
also universality of the soft susy breaking terms at the GUT scale 
in order to solve for those quantities too. This assumption can however
be relaxed \cite{Arnowitt} but still for small $\tan \beta$. 
It is thus natural to
try to find exact solutions for the Yukawa sector for any value of $\tan \beta$,
for comparable top and bottom Yukawa couplings, and also bringing in the
game the $\tau$ lepton Yukawa coupling as well in order to cope with
the case of $b-\tau$ unification \cite{btau}. Some attempts have been made 
(for instance in \cite{EL1}) to solve generally the top-bottom Yukawa system
which lead  to implicit solutions provided one neglects the
$U(1)_Y$ gauge coupling (see also \cite{EL2}). 

In the present paper we study some properties of the runnings of the Yukawa
couplings as dictated by the RGE's to one-loop. 
The first aim is to provide suitable expressions for the exact solutions,
which we call ``integrated forms''. Although these expressions do not appear
in closed forms, they are especially convenient because, if one insists on
making them explicit, they come out as continued integrated fractions 
the convergence of which can be kept under control. Although our main results
are {\sl a priori} valid for any gauge theory with an arbitrarily extended 
Yukawa sector ( including the special case of the standard model), 
we restrict most of the discussions and further illustrations
to the case of the MSSM.
We will give integrated forms,
valid for any number of Yukawa couplings, of the general explicit solutions
corresponding to the coupled renormalization group
equations which read in the MSSM \cite{falck}\\

\nn $\bullet$ Gauge Couplings [$g_i$ with $i=1,2,3$ and $n_g$ the generation 
number]: 
\beqn 
\label{coupjauge}
{\frac {dg_i}{dt}}={\frac {1}{32 \pi^2} b_i g_i^3 } \ \ {\rm with} \ \ 
b_1=-1-{\frac {10}{3}} n_{g} \ , \ b_2=5-2 n_{g} \ , \ b_3=9-2 n_{g}
\eeqn 
\nn $\bullet$ Yukawa Couplings [$i=1,2,3$ generations]: 
\begin{eqnarray}
 {\frac {dY_{u}^i}{dt}}&=&-{\frac {Y_{u}^i}{32 \pi^2}} \bigg[ 
 3 (Y_u^i)^2 +\! 3 \sum_{k=gen} (Y_u^k)^2+\!(Y_d^i)^2 - \bigg({\frac {13}{9}} 
g_1^2+\!3 g_2^2+\!\frac {16}{3}g_3^2 \bigg) \bigg] \label{yukup} \\
{\frac {dY_{d}^i}{dt}}&=&-{\frac {Y_{d}^i}{32 \pi^2}} \bigg[
{ 3 (Y_d^i)^2 +\! (Y_u^i)^2 +\! \sum_{k=gen} \{ 3 (Y_d^k)^2+\! (Y_l^k)^2 \}}  
-\bigg({ \frac {7}{9}} g_1^2+\!3 g_2^2+\!\frac {16}{3}g_3^2 \bigg) \bigg]
\label{yukdown} \\
{\frac {dY_{l}^i}{dt}}&=&-{\frac {Y_{l}^i}{32 \pi^2}} \bigg[ 
{ 3 (Y_l^i)^2 +\!\sum_{k=gen} \{ (Y_l^k)^2+\!3 (Y_d^k)^2\}}-3( g_1^2+\! g_2^2) \bigg]
\label{yuklep} 
\end{eqnarray}

Here the evolution parameter $t$ is defined by $t={\rm Log}(M_U^2/Q^2)$ where
$M_U$ denotes some initial scale. Note that since a gauge coupling unification
condition is not essential in the present study, we write the RGE equations
in terms of the low energy $SU(3)_c \times SU(2)_L \times U(1)_Y$ gauge 
couplings, respectively $g_3, g_2$ and $g_1$. Note also that we assume here,
and throughout the paper, flavour conserving (diagonal) Yukawa matrices.\\

The rest of the paper is organized as follows. In section 2 we recall
the known solution for large top Yukawa coupling and give the
integrated form of the general solution valid for any value of
the top and bottom Yukawas. We then generalize those integrated forms
to any number of Yukawa couplings, in particular to the
top-bottom-$\tau$ system. In section 3 we give a proof of the
convergence of these forms in both top--bottom and top--bottom--$\tau$ cases. 
Section 4 is devoted to the question of avoiding Landau poles in the Yukawa
runnings. There we give a generalization to the top--bottom case 
of some well known bounds, and establish necessary and sufficient conditions.
Preliminary applications and comments are made in section 5 and conclusions and
an outlook are given in section 6.
An appendix contains some detailed proofs and technical
material. 

\section{Integrated form of the Yukawa coupling RGE's}
\setcounter{equation}{0}
\subsection{Large top quark Yukawa solutions: a reminder} 
We are interested here in eqs.~(\ref{yukup}--\ref{yuklep}). They can be
treated independently of
 the rest of the system, especially
from the gauge couplings for which the running is determined a priori via 
eq.~(\ref{coupjauge}). [This is no more true at two--loop order where the 
gauge and Yukawa equations become highly interwound.] When all Yukawa couplings 
except $Y_t$ are neglected, eqs.~(\ref{yukdown}) and (\ref{yuklep}) become 
trivial while eq.~(\ref{yukup}) becomes of the Bernoulli type in the variable 
$y_t\equiv Y_t^2$ 
\begin{equation}
\frac{d}{dt} y_t = f_1(t) y_t + b y_t^2
\end{equation}
where 
\begin{eqnarray}
f_1(t)= \frac{1}{16 \pi^2}( \frac{16}{3} g_3^2 + 3 g_2^2 + 
\frac{13}{9} g_1^2 ) \ \ , \ \  b = -\frac{6}{16 \pi^2}  \label{f1} 
\end{eqnarray}
and is easily solved to give (\cite{diffeq},\cite{solutions}) 
\beqn
y_t(t)=\frac{y^0 E(t)}{1 -b y^0 \int_0^t E(t') dt'} \label{ytsol0}
\eeqn 
where 
\beqn  
E(t)= e^{\int_0^t f_1(t') dt'}
 \ \ {\rm and} \ \ y^0= Y_t^2(t=0) 
\eeqn
\subsection{top-bottom case}
In the more general case where both $Y_t$ and $Y_b$ are kept in the game, but 
neglecting all other Yukawa couplings, eqs.~(\ref{yukup},\ref{yukdown}) 
read, after the change of variables $y_t\equiv Y_t^2, y_b\equiv Y_b^2$,
\begin{eqnarray}
\frac{d}{dt} y_t= f_1(t) y_t + a y_b y_t + b y_t^2 \ \ , \ \ 
\frac{d}{dt} y_b= f_2(t) y_b + a y_b y_t + b y_b^2 \label{ytb} 
\end{eqnarray}
where $f_1(t)$ and $b$ are given in eqs.~(\ref{f1}) and 
\begin{eqnarray}
f_2(t)= \frac{1}{16 \pi^2}( \frac{16}{3} g_3^2 + 3 g_2^2 + 
\frac{7}{9} g_1^2 ) \ \ , \ \  a= -\frac{1}{16 \pi^2} 
\end{eqnarray}
As far as we know, the system eqs.~(\ref{ytb}) is not treated in standard
text books, and although it looks at first sight simple, we could not find 
a systematic way of relating it to a standard form
\footnote{The situation would be much simpler if $f_1(t) = f_2(t)$,  
in which case the equations can be solved by quadrature 
after some change of variables, leading though only to implicit solutions 
involving some hypergeometric functions \cite{EL1}}. It is also relatively easy 
to solve the system up to first order in $Y_b$ in the region $Y_t \gg Y_b$. 
This is already an improvement of the known solutions with $Y_b 
\sim 0$. It extends the numerical validity much further than $\tan \beta 
\simeq 10$.

More importantly, this approximate solution gives a valuable hint for 
the structure of a suitable integrated form which can then be found by sheer 
guess and will  be written down below. But first, the approximate
solution can be obtained in the form $ y_t(t) = \tilde{y}_t(t) + \delta(t)$
where $\tilde{y}_t(t)$ is given by eq.(\ref{ytsol0}) and linearizing
when necessary the equations, in the regime $y_b(t), |\delta(t)| \ll 1$. 
One then finds for $y_b$

\begin{equation}
y_b(t) =  \frac{y_b^0 E_{21}(t)}{1 - b y_b^0 \int_0^t E_{21}(t') dt'}
\label{approxyb}
\end{equation}

where 
\begin{eqnarray}
&&E_{21}(t) = \frac{ E_2(t)}{ (1 - b y_t^0 \int_0^t E_{1}(t') dt')^{a/b} } \\
&& \nonumber \\
&& E_i(t)= e^{\int_0^t f_i(t') dt'} \;\;\; i=1,2 \label{Ei}\\
\nonumber
\end{eqnarray}

and a slightly more complicated expression for $y_t$ given in appendix B.  
A little thinking then leads to the following form of exact ``solution'' we
were looking for:
  
\begin{eqnarray}
&&y_t(t)=\frac{y_t^0 E_{12}(t)}{1 - b y_t^0 \int_0^t E_{12}(t') dt'} \label{ytsol} \\
&& \nonumber \\
&&y_b(t)=\frac{y_b^0 E_{21}(t)}{1 - b y_b^0 \int_0^t E_{21}(t') dt'} \label{ybsol}
\end{eqnarray}
where
\begin{eqnarray}
&&E_{12}(t)= \frac{ E_1(t)}{ (1 - b y_b^0 \int_0^t E_{21}(t') dt')^{a/b}} 
\label{e12} \\
&& \nonumber \\
&&E_{21}(t)= \frac{ E_2(t)}{ (1 - b y_t^0 \int_0^t E_{12}(t') dt')^{a/b} } 
\label{e21}  \\
\nonumber
\end{eqnarray}
and $y_t^0\equiv Y_t^2(t=0), y_b^0\equiv Y_b^2(t=0)$ are any initial conditions.
The reader can easily check that the solutions (\ref{ytsol},\ref{ybsol}) satisfy
 {\sl exactly} Eqs.(\ref{ytb}) without any restriction or assumption
about the magnitudes of the Yukawa couplings, {\sl i.e.} for any value of 
$\tan \beta$.
They resemble formally eq.~(\ref{ytsol0}) of which they are a generalization.
Of course, although our solutions, $y_t, y_b$ are now explicit in terms
of $E_{12}$ and $E_{21}$, the latter are given only implicitly by 
Eqs.(\ref{e12}, \ref{e21} ), 
which appear as coupled nonlinear integral equations.
Therefore, the procedure is useful only if it provides us with a systematic
(and hopefully quick) way to solve these equations within a given accuracy.
It will be shown in sect. 3 that mere iterations achieve this goal.
In fact, such iterations correspond to the truncations of the ``continued 
integrated fractions'' which naturally emerge as {\sl formal} solutions
of Eqs.(\ref{e12}, \ref{e21} ), {\sl e.g.}:

\begin{eqnarray}
&&E_{12}(t)=\cfrac{E_{1}(t)}{( 1- b y_b^0 {\int_0^t}
           \cfrac{ E_{2}(t_1) dt_1}{( 1- b y_t^0 \int_0^{t_1}
            \cfrac{ E_{1}(t_2) dt_2}{( 1- b y_b^0 \int_0^{t_2}
             \cfrac{ E_{2}(t_3) dt_3}{( 1- b y_t^0 \int_0^{t_3} 
              \cfrac{E_1(t_4) dt_4}{\ddots}
              )^{a/b}}  )^{a/b}} )^{a/b}} )^{a/b}} \nonumber \\ 
\label{iterated}
\end{eqnarray}

\subsection{Arbitrary number of Yukawa fermions}
In fact the above solutions are easily generalized to include any
number of leptons and quarks. For instance, if one includes in the
game a third Yukawa coupling, then 
Eqs.(\ref{yukup}, \ref{yukdown}, \ref{yuklep}) take the following
form:
\begin{eqnarray}
\frac{d}{dt} y_1 &=& f_1(t) y_1 + a_{11} y_1^2 + a_{12} y_1 y_2 + a_{13} y_1 y_3 \nonumber \\
\frac{d}{dt} y_2 &=& f_2(t) y_2 + a_{22} y_2^2 + a_{21} y_2 y_1 + a_{23} y_2 y_3 \nonumber \\
\frac{d}{dt} y_3 &=& f_3(t) y_3 + a_{33} y_3^2 + a_{31} y_3 y_1 + a_{32} y_3 y_2 \nonumber \\
\end{eqnarray}

The exact solution reads:
\begin{eqnarray}
y_1 &=& \frac{y_1^0 u_1}{1 - a_{11} y_1^0 \int u_1} \nonumber \\
y_2 &=& \frac{y_2^0 u_2}{1 - a_{22} y_2^0 \int u_2} \nonumber \\
y_3 &=& \frac{y_3^0 u_3}{1 - a_{33} y_3^0 \int u_3} \nonumber \\
\label{threeyuk}
\end{eqnarray}

where $u_1, u_2$ and $u_3$ are defined through the implicit system

\begin{eqnarray}
u_1= \frac{E_1}{( 1- a_{22} y_2^0 \int u_2)^{a_{12}/a_{22}} 
(1- a_{33} y_3^0 \int u_3)^{a_{13}/a_{33}}} \nonumber \\
u_2= \frac{E_2}{( 1- a_{11} y_1^0 \int u_1)^{a_{21}/a_{11}} 
(1- a_{33} y_3^0 \int u_3)^{a_{23}/a_{33}}} \nonumber \\
u_3= \frac{E_3}{( 1- a_{11} y_1^0 \int u_1)^{a_{31}/a_{11}} 
(1- a_{22} y_2^0 \int u_2)^{a_{32}/a_{22}}} \nonumber \\
\label{threeEs}
\end{eqnarray}

and $\int u_j$ stands for $\int_0^t dt' u_j(t')$.

In the interesting case of top-bottom-$\tau$ system with $y_t\equiv  y_1$,
$y_b\equiv  y_2$ and  $y_\tau \equiv  y_3$ one has in the MSSM
\begin{eqnarray}
a_{11} = a_{22} = -\frac{6}{ 16 \pi^2} &\; ; \; &
a_{33}=-\frac{4}{16 \pi^2} \nonumber \\ 
\frac{a_{12}}{a_{22}} = \frac{a_{21}}{a_{11}} = \frac{1}{6} &\; ; \;&
\frac{a_{31}}{a_{11}} = \frac{a_{13}}{a_{33}} = 0 \nonumber \\
\frac{a_{23}}{a_{33}} = \frac{1}{4} &\; ; \;& 
\frac{a_{32}}{a_{22}} = \frac{1}{2} \nonumber  \\
f_3(t) = \frac{3}{16 \pi^2} (g_1^2 + g_2^2) 
&\; ; \;& E_3(t)= e^{\int_0^t f_3(t') dt'} \nonumber \\
\label{tbtau}
\end{eqnarray}

and $f_{1,2}(t), E_{1, 2}(t)$ as previously. It is interesting to note
that in this case $u_{\tau}$ and $u_t$ are directly related
via

\begin{equation}
\frac{u_{\tau}}{E_3} = \big ( \frac{u_t}{E_1} \big )^3 \label{special}
\end{equation}
a reflection of the fact that, in the MSSM, the running of $y_t$ and 
$y_{\tau}$ at one-loop order are mutually affected only indirectly through the 
running of $y_b$,
( $ a_{31} = a_{13} = 0 $) at variance with the non supersymmetric 
Standard Model (SM) case.\\

Finally, the extension to more than three Yukawa couplings will not play any 
role in the present paper. Nevertheless, we give it for the sake of 
completeness in appendix C.

\section{Proof of convergence}
\setcounter{equation}{0}
\subsection{The top-bottom case:} 

In this section we make a mathematical digression to study
some useful properties of our solutions.
Even though equations (\ref{e12}, \ref{e21}) give $E_{i j}$
only implicitly, they enjoy the property
of defining a {\sl contraction mapping}. This is about all what
one needs to give a rigorous proof for the existence and
uniqueness of the $E_{i j}$, and thus of the existence
and uniqueness of the solutions given in Eq.(\ref{ytsol}, \ref{ybsol}). 
This proof will also be of practical use. It provides us with a criterion
for the convergence of the truncated forms of Eq.(\ref{iterated}) towards
the exact solution, and the rate of this convergence can be controlled so
that a very good approximation will be obtained with a few (or even just one)
iterations.\\

For the sake of completeness, we recall here in simple terms
the conditions required for a contraction mapping, and then prove
that they are indeed satisfied in our case. Let us define
\begin{eqnarray}
&& U_1(t) =\frac{E_{12}(t)}{E_1(t)} \\
&& U_2(t) = \frac{E_{21}(t)}{E_2(t)} \\
\nonumber
\end{eqnarray}
and think of $U_1$ and $U_2$ as forming a vector 

\begin{equation}
\vec{U}(t)= \left(
\begin{array}{l}
U_1(t) \\
U_2(t) 
\end{array} \right)
\label{Uvect}
\end{equation}
in some space ${\cal E}_T$ where the evolution parameter $t$ remains in
the interval $ 0\leq t \leq T$ for a given value of $T$. Then
Eqs.(\ref{e12}, \ref{e21}) restrict the $U_i's$ to the positive
region $ U_i(t) \geq 0$ (provided one stays far from the Landau poles),
and, furthermore, define a mapping in this region, 
$ {\cal A}: \vec{U} \mapsto \vec{U}'$
through

\begin{eqnarray}
U_1'(t) = \frac{1}{( 1 - b y_2^0 \int_0^t E_2(t') U_2(t') dt')^{a/b}} \nonumber
\\
U_2'(t) = \frac{1}{( 1 - b y_1^0 \int_0^t E_1(t') U_1(t') dt')^{a/b}} \nonumber \\
\label{mapping}
\end{eqnarray}

The idea now is to show that the mapping ${\cal A}$ shrinks uniformly,
at each iteration, 
the ``distance'' between any two vectors in ${\cal E}_T$ 
(subject to the condition $ U_i(t) \geq 0$). 
More precisely we will prove that there exists a positive constant number 
$K_T < 1$
such that the following inequality is satisfied

\begin{equation}
\parallel \vec{U'} - \vec{V'}\parallel \leq K_T 
\parallel \vec{U} - \vec{V}\parallel
\label{contraction}
\end{equation}  
for any pair of vectors $( \vec{U}, \vec{V})$ belonging to
${\cal E}_T$ and satisfying $ U_i, V_i \geq 0$ ( $i=1 ,2 $).
Here $\parallel . \parallel$ is defined by
\begin{equation}
\parallel \vec{U} \parallel = \max \{ \sup_{0 \leq t \leq T} \mid U_1(t) \mid,
\sup_{0 \leq t\leq T} \mid U_2(t)\mid \}
\label{normdef}
\end{equation}

Then, according to the ``contraction mapping principle'', the existence
of the (unique) solution of Eqs.(\ref{e12}, \ref{e21}) in ${\cal E}_T$ is
guaranteed, and the $n^{th}$ iteration ${\cal A}^n \vec{U}$ approaches this
solution at least as fast as $K_T^n$.  

To prove Eq.(\ref{contraction}) one writes the following sequence of
inequalities

\begin{eqnarray}
\mid U_i'(t) - V_i'(t) \mid &\leq&
\frac{y_j^0 \int_0^t dt' E_j(t')  \mid U_j(t') - V_j(t') \mid}{
[ 16 \pi^2 + 6 y_j^0 \min\{\int_0^t dt' E_j(t') U_j(t'), 
\int_0^t dt' E_j(t') V_j(t')\} ]^{7/6} } \nonumber \\
&\leq & \frac{y_j^0}{16 \pi^2}  \sup_{0 \leq \tau \leq T}\mid U_j(\tau) - V_j(\tau) \mid 
\int_0^T dt' E_j(t') \nonumber \\
&\leq & \frac{y_j^0}{16 \pi^2}  \parallel \vec{U} - \vec{V} \parallel \int_0^T 
dt' E_j(t')
\nonumber \\
\label{ineq2} 
\end{eqnarray}  

valid for $i\neq j$ with $i, j= 1, 2$ with $y_1 \equiv y_t, y_2 \equiv y_b$
 and where we plugged the actual
values of the coefficients $a, b$ in Eq.(\ref{ytb}). The first inequality
in (\ref{ineq2}) is derived from Eqs.(\ref{mapping}) and from the inequality,
 
\begin{equation}
\mid (\frac{1}{1 + \alpha})^c - (\frac{1}{1 + \beta})^c \mid \leq c
\frac{\mid \alpha - \beta \mid}{( 1 + \min\{\alpha, \beta\})^{c + 1}},
\label{ineq1}
\end{equation}

valid for any $\alpha, \beta$, larger than $-1$ and $c  > 0$ (here $c = 1/6$)

the second from the positivity of $ U_i, V_i, E_i$ and $y^0_i$,
and the third from the definition (\ref{normdef}). 
Eq.(\ref{ineq2}) immediately leads to Eq.(\ref{contraction})
with
\begin{equation}
K_T  =\frac{1}{16 \pi^2} 
max\{y_t^0 \int_0^T dt E_1(t),
y_b^0 \int_0^T dt E_2(t)\}
\end{equation} 

The convergence condition $K_T < 1$ is easily met even when $T$ is large enough
to encompass the whole evolution range from the GUT scale to the $M_Z$
scale. For instance, if $T \approx 66$ and $\alpha_{GUT}^{-1} \approx 25$,
one needs $ Y^0_t, Y^0_b \lsim O(\pi) $ 
where the $Y^0_i$ are the GUT scale values of the Yukawa couplings, to ensure
convergence. These conditions are naturally met within the perturbative
regime. We should stress, however, that these are only sufficient conditions.

\subsection{The top-bottom-$\tau$ case:}
As we said previously the exact solutions (\ref{threeyuk},
\ref{threeEs}) to the generalized  
equations with three Yukawa couplings or more, is a direct generalization
of eqs.(\ref{ytsol}, \ref{ybsol}). Similarly the proof of convergence goes 
essentially along the same lines as in the previous section, when
supplemented with the inequality 


\begin{eqnarray}
&&\mid \frac{1}{(1 + \alpha_1)^a} \frac{1}{(1 + \beta_1)^b} -
\frac{1}{(1 + \alpha_2)^a} \frac{1}{(1 + \beta_2)^b} \mid \leq \nonumber \\
&& \frac{a \mid \alpha_1 - \alpha_2 \mid}
{( 1 + \min\{\alpha_1, \alpha_2\})^{a + 1} }
+ \frac{b \mid \beta_1 - \beta_2 \mid} 
{( 1 + \min\{\beta_1, \beta_2\})^{b + 1} } \nonumber \\
\label{ineq3}
\end{eqnarray}

valid for any $a,b, \alpha_i, \beta_i > 0 $. 
Defining a mapping through eqs.(\ref{threeEs}), one finds the convergence
criterion
\begin{equation}
K_T = \max_{(i j k)circ.perm. of (1 2 3)} \{ -a_{i j} y_j^0 \int_0^T E_j
-a_{i k } y_k^0 \int_0^T E_k \} < 1
\end{equation} 

In the top-bottom-$\tau$ case it reads

\begin{eqnarray}
K_T = \frac{1}{16 \pi^2} \max \{ y_b^0 \int_0^T E_2 \; ;\; y_t^0 \int_0^T E_1 +
y_\tau^0 \int_0^T E_3 \; ; \; 3 y_b^0 \int_0^T E_2 \} = && \nonumber \\
\frac{1}{16 \pi^2} \max \{y_t^0 \int_0^T E_1 +
y_\tau^0 \int_0^T E_3 \; ; \; 3 y_b^0 \int_0^T E_2 \} < 1 && \nonumber \\
\end{eqnarray}
We see that the sufficient convergence criterion can become more severe in this case by about a factor 3 in the regime $y^0_t \sim y^0_b \sim y^0_{\tau}$.

 
\section{Avoiding Landau poles}
\setcounter{equation}{0}
One question that can be clearly answered with the knowledge of 
analytical solutions is how to determine the conditions which guarantee that
the values of the Yukawa couplings at some low-energy scale,
say the electroweak scale, remain consistent with a Landau ``pole" free theory
up to a given high energy scale, typically a grand unification scale. 
The answer would be trivial if one starts from the high energy scale
Yukawa coupling and runs down. Indeed in this case, it is clear from
the general form of the solutions, eqs.(\ref{ytsol} , \ref{ybsol})
and the fact that $b <0, y^0 (\equiv {Y^0}^2) \geq 0$ and $ E(t)_{i j} \geq 0$,
that one does not hit a Landau pole all the way down below the initial 
scale\footnote{Note that here we are only interested in poles which can occur
explicitly in the Yukawa couplings. In practice one stays anyway far from
gauge coupling Landau poles, given the running range relevant to our discussion.}.
The situation is more complicated if one starts from some Yukawa coupling
values at a low scale and tries to run upwards to determine the corresponding
values at a GUT scale. This is a phenomenologically typical
situation if a model-independent reconstruction of the fundamental parameters
is to be carried out, starting from the experimental input.\\ 

Moreover since in the vicinity of such Landau poles the Yukawa couplings
become very large, the conditions for avoiding these poles correspond
in some cases to effectively attractive fixed points, such as the   
celebrated relation
\begin{equation}
m_{top} \approx ( 190 - 200 GeV) \sin \beta
\end{equation}
in the MSSM valid when all Yukawa couplings are neglected in comparison to the 
that of the top quark \cite{Alvarez}. When the Yukawa couplings are of
comparable values such a correspondence becomes more involved, as can be
seen for instance from the dependence on $y_t^0$ and $y_b^0$ in (\ref{iterated}) [See also the discussion in section {\bf 5.2}].

To illustrate the case, we start first with the solution for
small $\tan \beta$, i.e. $Y^0_t \gg Y^0_b \sim 0$, given in eq.(\ref{ytsol0}).
Writing it in the form

\begin{equation}
y^0 = \frac{y_t(t)}{ E(t) + b y_t(t) \int_0^t E(t') dt'}
\end{equation}
one sees immediately that $y_t(t)$ should satisfy

\begin{equation}
y_t(t) < -\frac{E(t)}{b \int_0^t E(t')dt'}
\label{constraint1}
\end{equation}
for any value of $t >0$ in order to be consistent with the positivity of
$y^0$ and $ y_t(t)$. Ensuring this positivity avoids automatically
the Landau pole ( we rely here on the
fact that $b<0$, see eq.(2.2) .) Thus contrary to the initial value
at a high scale $t=0$, values of $y_t$ can not be arbitrarily chosen at lower 
scales, the maximum allowed in this case being simply given by 
eq.(\ref{constraint1}). 
With this in mind, it is straightforward to eliminate completely the dependence
on $y^0$ in the running to get

\begin{equation}
y_t(t) = \frac{ y_t(t_0) E(t; t_0)}{ 1 - b y_t(t_0) \int_{t_0}^t E(t'; t_0) dt'}
\end{equation}

where 
\begin{equation}
E(t;t_0) \equiv \frac{E(t)}{E(t_0)}
\end{equation}

Here $y_t(t_0)$ is any initial value satisfying eq.(\ref{constraint1})
for $t = t_0$.\\

In the more general case, when $y_b$ is not neglected, one can also write,

\begin{eqnarray}
y_b(t) = \frac{ y_b(t_0) E_{21}(t; t_0)}{ 1 - b y_b(t_0) \int_{t_0}^t E_{21}(t'; t_0) dt'}
\nonumber \\ 
y_t(t) = \frac{ y_t(t_0) E_{12}(t; t_0)}{ 1 - b y_t(t_0) \int_{t_0}^t E_{12}(t'; t_0) dt'}
\nonumber\\
\label{y0sol}
\end{eqnarray}

with 
\begin{eqnarray}
E_{12}(t;t_0) = \frac{E_1(t;t_0)}{( 1 - b y_b(t_0) \int_{t_0}^{t} 
E_{21}(t';t_0) dt')^{a/b}} \nonumber \\
E_{21}(t;t_0) = \frac{E_2(t;t_0)}{( 1 - b y_t(t_0)  \int_{t_0}^{t} 
E_{12}(t';t_0) dt' )^{a/b}}
 \nonumber \\
\label{e1221}
\end{eqnarray}

where 

\begin{equation}
E_j(t;t_0) = e^{-\int_t^{t_0}dt' f_j(t')} = \frac{E_j(t)}{E_j(t_0)}
\end{equation}
Note that here $t_0 \geq t$ corresponds to an initial energy scale
{\sl lower} than the running scale corresponding to $t$. 
In this case, however, it is no more possible to determine easily  
a sufficient and necessary condition (if any) on $y_b(t_0)$ and
$y_t(t_0)$ to avoid the Landau pole. 
Indeed the sufficient and necessary conditions for a non singular running
in the interval $[T, t_0]$ read
\begin{eqnarray}
&& 1 - b y_b(t_0) \int_{t_0}^{T} 
E_{21}(t';t_0) dt' > 0  \label{necsuf1}\\
&& 1 - b y_t(t_0)  \int_{t_0}^{T} 
E_{12}(t';t_0) dt'  > 0 \label{necsuf2}
\end{eqnarray}
as easily seen from eqs.(\ref{y0sol}). However, $E_{12}$ and $E_{21}$ depend
themselves on $y_t^0$ and $y_b^0$ so that the above conditions are highly 
implicit in $y_t^0$ and $y_b^0$. 

Instead, one can immediately determine some {\sl necessary} conditions,
and, with some extra work, also some {\sl sufficient} ones.
We list these conditions below and refer the reader to the appendix for
the detailed proofs.\\

{\sl The necessary conditions:} one can write a tower of pair of inequalities,
each pair being a necessary condition weaker than the subsequent one,

\begin{eqnarray}
&&\left\{\begin{array}{l}
y_t(t_0) < \cfrac{E_1(t_0)}{ |b| \int_T^{t_0} dt E_1(t)} \label{necess1}\\
{ }  \\ {} 
y_b(t_0) < \cfrac{E_2(t_0)}{ |b| \int_T^{t_0} dt E_2(t)} \label{necess2} 
\end{array}\right. \\
{ }\nonumber \\ {} \nonumber
{ }\\ {} 
&&\left\{\begin{array}{l}
y_t(t_0) < \cfrac{E_1(t_0)}{ |b| \int_T^{t_0} \cfrac{dt_1 E_1(t_1)}{
(1 - |b| y_b(t_0) \int_{t_1}^{t_0} E_2(t_2;t_0) dt_2)^{a/b} }} \label{necess11}\\
{ }\\ {}
y_b(t_0) < \cfrac{E_2(t_0)}{ |b| \int_T^{t_0} \cfrac{ dt_1 E_2(t_1)}{
(1 - |b| y_t(t_0) \int_{t_1}^{t_0} E_1(t_2;t_0) dt_2)^{a/b} }}\label{necess21} 
\end{array}\right. 
{ }\\ {} \nonumber
{ }\\ {}
&& \;\;\;\;\;\;\;\;\;\;\;\;\;\;\;\;\vdots  \nonumber
{ }\\ {}  \nonumber
{ }\\ {}
&& \;\;\;\;\;\;\;\;\;\;\;\;\;\;\;\;\vdots  \nonumber 
{ }\\ {}  \nonumber
{ }\\ {}
&&\left\{\begin{array}{l}
y_t(t_0) < \cfrac{1}{ |b| \int_T^{t_0} dt E_{12}(t;t_0)} \label{necess1inf}\\
{ }  \\ {} 
y_b(t_0) < \cfrac{1}{ |b| \int_T^{t_0} dt E_{21}(t;t_0)} \label{necess2inf} 
\end{array}\right. \\ \nonumber
\end{eqnarray}
The limit of this tower of inequalities (assuming of course that the iteration
converges) is precisely the necessary and sufficient condition
eqs.(\ref{necsuf1}, \ref{necsuf2}), as can be seen from eqs.(\ref{e1221}) 
when written in a form similar to eq.(\ref{iterated}).\\

{\sl The sufficient conditions:}
\begin{equation}
y_t(t_0) < \frac{1}{c ( 1 + 1/c)^{1 + c}}\frac{E_1(t_0)}{ |b| \int_T^{t_0} dt E_1(t)} \label{suffice1}
\end{equation}

\begin{equation}
y_b(t_0) < \frac{1}{c ( 1 + 1/c)^{1 + c}}\frac{E_2(t_0)}{ |b| \int_T^{t_0} dt E_2(t)} \label{suffice2}
\end{equation}
when $ c = a/b$.
It is interesting to note that the above conditions have basically the same 
form as (\ref{necess2}), apart from the factor
 $\frac{1}{c ( 1 + 1/c)^{1 + c}}$ ($\sim 0.62$ in the MSSM.)

\section{Preliminary applications and comments}

The aim of this section is to illustrate briefly through some examples the 
possible
use of the integrated forms and make contact with the existing studies
and approximations. It is, however, obviously not meant to be exhaustive
nor refined from the phenomenological point of view (for instance no 
threshold effects or higher loop
effects are included), as this would deserve a separate analysis by itself. 
Let us also keep in mind that all the functions $E_i$ which enter the solutions
are analytically known in terms of the initial gauge coupling values as can
been seen from eq.(\ref{coupjauge}) and its solutions. 
 
\subsection{Landau pole free bounds:}
The necessary and sufficient bounds found in the previous section are
exact generalization of the one in \cite{Hill} initially derived in the 
regime of large $y_t ( \gg y_b)$.  
The necessary bounds (\ref{necess2}) which would be also
sufficient in the limit $ a \to 0$ ( where they become identical to
(\ref{suffice1}, \ref{suffice2}) ) give a first estimate of the
allowed values for $y_t, y_b$ at low energy, restricting them to a rectangle

\begin{equation}
y_t^{MSSM}(EW) < \overline{y}_t\;, \;\; y_b^{MSSM}(EW) < \overline{y}_b 
\label{applibounds}
\end{equation}

where $\overline{y}_t$ and $\overline{y}_b$ depend on the gauge couplings and 
are easily determined numerically, for instance in terms of a grand unified scale, the value of the gauge 
couplings at that scale, and some low energy electroweak scale.
 

For instance, one has roughly [neglecting the $g_1$ and $g_2$ couplings],

\begin{equation}
\overline{y_{t,b}} = \frac{7}{18} \frac{g_3^4(EW)}{ 1 - (\frac{g_3^2(EW)}{g_3^2(GUT)})^{-7/9}}
\end{equation}

We stress that the necessary bounds in eq.(\ref{applibounds})
involve no approximation whatsoever. One can readily turn them into
constraints on the $\tan \beta$ parameter at some electroweak scale:

\begin{equation}
\frac{(\overline{m}_{top} \overline{g}_2)^2}{(\sqrt{2 \overline{y}_t} \overline{M}_W)^2 -
(\overline{m}_{top} \overline{g}_2)^2 } < \tan^2 \beta <
\frac{(\sqrt{2 \overline{y}_b} \overline{M}_W)^2 - (\overline{m}_{b} \overline{g}_2)^2 }{
(\overline{m}_{b} \overline{g}_2)^2} \label{tanbound}
\end{equation}

The bars indicate that the masses and gauge couplings are running quantities
at the chosen low energy scale.

Going now to the improved bounds (\ref{necess21}) one can reduce further
the allowed range for the Yukawa couplings. These bounds 
do not allow in their general form an easy analytic determination
of the allowed regions. To get a feeling about these regions,
let us illustrate the case in two different approximations:

\begin{itemize}
\item[{\sl i)}] $E_1 = E_2 \equiv E$, {\sl i.e.} neglect the difference 
$f_1 - f_2 = g_1^2/24\pi^2$
\item[{\sl ii) }] assume $y_b(t_0), y_t(t_0)$ sufficiently small for a first
order expansion to be legitimate
\end{itemize}

{\sl i)} In this case one integral can be performed exactly in 
(\ref{necess21}), leading to

\begin{eqnarray}
&&\left\{\begin{array}{l}
y_t(t_0) < \cfrac{y_b(t_0) ( 1 - \frac{a}{b} )}{
1 - ( 1 - |b| y_b(t_0) \int_T^{t_0} dt_2 E(t_2; t_0))^{1 - a/b}} \label{necess1approx}\\
{ }  \\ {} 
y_b(t_0) < \cfrac{y_t(t_0) ( 1 - \frac{a}{b} )}{
1 - ( 1 - |b| y_t(t_0) \int_T^{t_0} dt_2 E(t_2; t_0))^{1 - a/b}}\label{necess2approx} 
\end{array}\right. \\
\nonumber
\end{eqnarray}

The domain defined by (\ref{necess2approx})
lies within the rectangle (\ref{applibounds}) and is controlled by the
relative strength of $a$ and $b$. For instance, the delimiting
curves start off at the points 
$(\overline{y}_b, 0)$ and $(0, \overline{y}_t)$ with
first derivatives equal to $-\frac{a}{2 b}$ ( $-\frac{1}{12}$ in the MSSM
and $-\frac{1}{6}$ in the SM), to be compared with $0$ in the 
rectangular approximation. Moreover, the exact fixed line solution
$y_t= y_b$ leads to the constraint 
\begin{equation}
y_t(t_0) = y_b(t_0) < \frac{E(t_0)}{ |a+b| \int_T^{t_0} dt E(t)}
\end{equation}

to be compared with the rectangular approximation (\ref{applibounds})
 where $|a + b|$ is
replaced by $|b|$, (a $14 \%$ effect in the MSSM and a $25 \%$ effect
in the SM.) These considerations give a qualitative guideline of the reduction
of the allowed domain.

{\sl ii)} In this case one gets a linearized approximation of the domain
[keeping though the effect of the $U(1)_Y$ coupling], 
in the following form, 

\begin{eqnarray}
&&\left\{\begin{array}{l}
|b| (\int E_1)^2 y_t( t_0) + |a| (\int E_1 \,\int E_2) y_b(t_0) < \int E_1
\\
{ }  \\ {}
|b| (\int E_2)^2 y_b( t_0) + |a| (\int E_2 \,\int E_1) y_t(t_0) < \int E_1
\end{array}\right. \\
\nonumber
\end{eqnarray}

where the $E_i$'s here are normalized to $E_i(t_0)$, 
$\int ... \equiv \int_T^{t_0} ...dt_1$ and \newline
$\int ... \int ... \equiv \int_T^{t_0} ... dt_1 \int_{t_1}^{t_0} ... dt_2$.

In this approximation the necessary domain is delimitated by two straight lines with scale
dependent slopes. Again, one can translate these conditions into bounds
on $\tan \beta$ at some effective electroweak scale.

\subsection{$y_t-y_b-g_3$ approximation, fixed points and quasi-fixed line:}

In the approximation where $E_1(t)= E_2(t)\equiv E(t)$, 
and assuming that the initial values $y_t^0, y_b^0$ are small enough
so that one iteration in the form of eqs.(\ref{e12},\ref{e21}) is a good 
approximation for the $E_{ij}(t)$, an easy integration yields

\begin{eqnarray}
y_t(t) = \cfrac{y_t^0 E(t)}{(1 - b y_b^0 \int E)^{a/b}} \cfrac{1}{
 \big [1 + \frac{y_t^0}{y_b^0} \frac{b}{b-a} ( (1 - b y_b^0 \int E)^{1-a/b} -1)
\big ]} \\
y_b(t) = \cfrac{y_b^0 E(t)}{(1 - b y_t^0 \int E)^{a/b}} \cfrac{1}{
 \big [1 + \frac{y_b^0}{y_t^0} \frac{b}{b-a} ( (1 - b y_t^0 \int E)^{1-a/b} -1)
\big ]}
\end{eqnarray}

provided of course all other Yukawas are put to zero. Note that at 
this level of approximations the above solutions depend just on one integral,
namely $ \int_0^t E$. If one goes further and neglects the $SU_L(2)$ gauge
coupling then  $ \int_0^t E$ can be computed explicitly, and one obtains,


\begin{eqnarray}
\rho_t(X)= \rho_t^0\cfrac{X}{\big[ 1 + \alpha \rho_b^0( X - 1) 
     \big ]^{c}} \;\cfrac{1}{\big [ 1 + \frac{\rho_t^0}{\rho_b^0}\frac{1}{1-c}
( \;( 1 + \alpha \rho_b^0 ( X - 1)\;)^{1 - c} - 1 ) \big ]}  
\label{approxg31} \\
\rho_b(X)= \rho_b^0\cfrac{X}{\big[ 1 + \alpha \rho_t^0( X - 1) 
     \big ]^{c}} \;\cfrac{1}{\big [ 1 + \frac{\rho_b^0}{\rho_t^0}\frac{1}{1-c}
( \;( 1 + \alpha \rho_t^0 ( X - 1)\;)^{1 - c} - 1 ) \big ]}  \label{approxg32} \\
\nonumber
\end{eqnarray}

where now we use the reduced variables
\begin{equation}
\rho_t \equiv \frac{y_t(t)}{g_3^2(t)} \;,\;\;
\rho_b \equiv \frac{y_b(t)}{g_3^2(t)} 
\end{equation}
and
\begin{equation}
X\equiv E(t)^{7/16} = \big (\frac{g_3^2(t)}{{g^0_3}^2} \big)^{7/9}
\end{equation}

(Note that $\alpha = 18/7, c=a/b=1/6$ in the MSSM and 
$\alpha = 9/2, c=1/3$ in the SM.)
The approximate solutions (\ref{approxg31}, \ref{approxg32}) allow one
to retrieve the well-known infrared fixed points in the $(\rho_t, \rho_b)$ 
plane, \cite{schrempp}. 
For instance, in the MSSM, the fixed point $(\rho_t=\frac{7}{18}, \rho_b=0)$
( respectively $(\rho_t=0, \rho_b=\frac{7}{18})$ ) is obtained
by looking at the limiting behaviour of $\rho_t(X)$ when $\rho_b^0 \to 0$
(resp. of $\rho_b(X)$ when $\rho_t^0 \to 0$. On the other hand, 
the IR (attractive) fixed point $(\rho_t=\frac{1}{3}, \rho_b=\frac{1}{3})$
is obtained by expanding $\rho_t(X)$ and $ \rho_b(X)$ simultaneously
for small $\rho_t^0$ and $\rho_b^0$. It should come as no surprise that
those exact fixed points are obtained only in this limit since the
solutions (\ref{approxg31}, \ref{approxg32}) become exact only in this limit.
[see also a related comment at the end of this section.] 

The exact fixed line $\rho_t=\rho_b$ is trivially obtained from
(\ref{approxg31}, \ref{approxg32}), {\sl i.e.} starting from 
$\rho_t^0=\rho_b^0$ the two reduced Yukawas remain equal at any other scale.
Perhaps more interesting is to ask whether one can determine analytically
the other exact or effective IR fixed lines. An IR attractive effective fixed
line of the form 

\begin{equation} \rho_t + \rho_b = \frac{2}{3} \label{flsch}
\end{equation}

 was found in \cite{schrempp2} for the MSSM. 

Starting from eqs.(\ref{approxg31}, \ref{approxg32}) one can actually 
improve on this effective fixed line in the following way. 
For small $\rho_t^0, \rho_b^0$ one obtains the integral line

\begin{equation}
\alpha ( 1-c) (\rho_b^0 - \rho_t^0) \rho_t \rho_b + 
\rho_t^0 ( 1 - \alpha (\rho_b^0 + c \rho_t^0)) \rho_b - 
\rho_b^0 ( 1 - \alpha (\rho_t^0 + c \rho_b^0)) \rho_t = 0
\end{equation}

If we require this integral line to go through the exact fixed point
$\rho_t = \rho_b= \frac{1}{\alpha (1 +c)}$ then 
a fixed line is obtained in the form

 

\begin{equation}
\frac{\rho_t + \rho_b}{ \rho_t \rho_b} = 2 ( 1+ c) \alpha \label{fl} 
\end{equation}

to first order in an expansion around the fixed point 
$\rho_t = \rho_b= \frac{1}{\alpha (1 +c)}$.

That is 

\begin{equation}
\frac{\rho_t + \rho_b}{ \rho_t \rho_b} = 6  \label{flMSSM}
\end{equation}

in the case of the MSSM, which constitutes an improved effective IR fixed line  
beyond Eq.(\ref{flsch}).
One can even reasonably expect Eq.(\ref{flMSSM}), and more
generally Eq.(\ref{fl}), to be exact in the regime under consideration. Indeed, 
for instance in the case of the SM, the effective fixed line (\ref{fl})
reads
  
\begin{equation}
\frac{\rho_t + \rho_b}{ \rho_t \rho_b} =12  \label{flSM}
\end{equation}

On the other hand, an exact IR fixed line is known in this case 
(see \cite{schrempp}) and is the sum of two terms, one of which coincides
precisely with (\ref{flSM}), the other being vanishing to first order in
the deviation, $\delta$, around the fixed point, that is for 
$\rho_t= 1/6 + \delta, \rho_b= 1/6 - \delta$.\\

Let us end this section by noting further possible applications  of the
integrated forms. As we mentioned before, if one starts from 
(\ref{approxg31}, \ref{approxg32}),  
one retrieves the exact fixed point $\rho_t=\rho_b= 1/(\alpha (1 + c))$ only in 
the region of small $\rho_t^0, \rho_b^0$. Moreover, in the deep infrared region ($X \to \infty$)
one obtains from Eqs.(\ref{approxg31}, \ref{approxg32})
$\rho_t=\rho_b= (1 - c)/\alpha$ irrespective of the initial values $\rho_t^0, \rho_b^0$.
The IR attraction to this point is of course an artifact of the approximate solutions.
In fact, one can resum exactly the integrated forms (\ref{e12}, \ref{e21}) in 
the limit
$X \to \infty$ and obtain  $\rho_t=\rho_b= 1/(\alpha (1 + c))$ as the attractive
IR fixed point, independently of the initial values $\rho_t^0, \rho_b^0$.\\

Other regimes can be also looked at (for instance $\rho_t^0 \ll \rho_b^0$ or $\rho_b^0 \ll \rho_t^0$)
for which approximate analytical expressions for the integrated forms can be obtained up
to three iterations, thus improving on (\ref{approxg31}, \ref{approxg32}). This allows to
tackle the form of the fixed line in such regimes. We do not dwell further on these aspects here.

\subsection{constraints in the $y_t-y_b-y_{\tau}$ and all gauge couplings case:}
In this section we illustrate the use of the solutions in the 
$t-b-\tau$ system only to derive inequalities 
which correlate the three fermion running masses in terms of initial
values for the Yukawa couplings. Starting from equations (\ref{threeyuk}-
\ref{tbtau}) and using the fact that
$(1 + |\alpha| y_i^0 \int u_i)^{|\beta|} \geq 1$ for all 
expressions of that form appearing in (\ref{threeyuk}, \ref{threeEs})
one writes immediately the following (necessary) inequalities:  

\begin{equation}
\overline{m}_t \leq Y_t^0 (2 \sqrt{2}G_F)^{-1/2} \sqrt{E_1} 
\sin \beta \label{ineqt}
\end{equation}

\begin{equation}
\overline{m}_i \leq Y_i^0 (2 \sqrt{2}G_F)^{-1/2} \sqrt{E_i} 
\cos \beta \label{ineqbtau}
\end{equation}

where $i= 2 (b), 3 ( \tau)$ and recalling that the $y_i's$ are squares of the 
Yukawa couplings $Y_i$. $Y^0_i$ denote the values of these couplings at some 
high energy scale and the bar denotes running quantities at the 
electroweak scale. The $E_i's$ are as defined in eqs.(\ref{Ei}, \ref{tbtau}).
These inequalities give boundary conditions on the initial values of
the Yukawa couplings, necessary to retrieve the correct "physical" fermion
masses.  For instance one immediately sees from (\ref{ineqbtau}) that large 
$\tan \beta$ necessitates large initial values for the bottom and $\tau$ 
Yukawa couplings. 

Moreover, relying systematically on the fact that
 $(1 + |\alpha| y_i^0 \int u_i)^{|\beta|} \geq 1$ for $i = 1,2,3$, one can  
derive the following optimal rigorous inequalities for more involved
combinations




\begin{eqnarray}
\frac{ y_b^{18}}{ y_t^3 y_{\tau}^{35}} &\geq& \frac{ (y_b^0)^{18}}{ (y_t^0)^3 
(y_{\tau}^0)^{35}}
\frac{ E_2^{18}}{ E_1^3 E_3^{35}} \\
\frac{ y_t^2 y_{\tau}^3}{ y_b^{12} } &\geq& 
\frac{ (y_t^0)^2 (y_{\tau}^0)^3}{ (y_b^0)^{12} }\frac{E_1^2 E_3^3}{ E_2^{12}} \\
\frac{ y_b^4}{ y_t^{21} y_{\tau} } &\geq& \frac{ (y_b^0)^4}{ (y_t^0)^{21} 
y_{\tau}^0}
\frac{ E_2^4}{ E_1^{21} E_3} \\
\nonumber
\end{eqnarray}

which can be readily translated into inequalities involving the running
quark masses, $\tan \beta$ and the three
gauge couplings (all taken at the electroweak scale), as well as the values
of the three Yukawa couplings at some initial scale

\begin{eqnarray}
\frac{ \overline{m_b}^{18}}{ \overline{m_t}^3 \overline{m_{\tau}}^{35} } &\geq&
\frac{ (Y_b^0)^{18}} {(Y_t^0)^3 (Y_{\tau}^0)^{35}} 
\frac{E_2^9}{ \sqrt{ E_1^3 E_3^{35}}} 
\frac{ (2 \sqrt{2}G_F)^{10}}{\sin^3 \beta \cos^{17} \beta} \label{ineqbetter1}\\
\frac{ \overline{m_t}^2 \overline{m_{\tau}}^3}{ \overline{m_b}^{12}} &\geq&
\frac{ (Y_t^0)^2 (Y_{\tau}^0)^3}{ (Y_b^0)^{12} } \sqrt{E_3^3} \frac{E_1}{E_2^6} \frac{\tan^2 \beta}{ \cos^7 \beta} (2 \sqrt{2} G_F)^{7/2} 
\label{ineqbetter2} \\
\frac{ \overline{m_b}^4}{ \overline{m_t}^{21} \overline{m_{\tau}} } &\geq&
\frac{ (Y_b^0)^4} {(Y_t^0)^{21} Y_{\tau}^0} 
\frac{E_2^2}{ \sqrt{ E_1^{21} E_3}} 
\frac{ (2 \sqrt{2}G_F)^9}{\tan^3 \beta \sin^{18} \beta} 
\label{ineqbetter3} \\
\nonumber
\end{eqnarray}

These inequalities express general necessary conditions which delineate the
physically allowed regions for the initial values of the three Yukawa couplings,
{\sl i.e.} consistent with the values of the physical top, bottom and $\tau$ masses, 
 prior to any model assumption\footnote{In a more refined treatment, one should of course
correct for the difference between the running and the pole masses}.
[Note also that (\ref{ineqt}, \ref{ineqbtau}) are already contained in 
(\ref{ineqbetter1} - \ref{ineqbetter3}).] Finally, had we neglected
the $\tau$ Yukawa coupling, the necessary inequalities involving the top and 
bottom running masses would have read

\begin{eqnarray}
\frac{ \overline{m_t}}{ \overline{m_b}^6 } &\geq&
\frac{ Y_t^0} {(Y_b^0)^6 } \frac{ \sqrt{ E_1}}{E_2^9} \frac{ \tan \beta}{ \cos^5 \beta} 
 (2 \sqrt{2}G_F)^{5/2} 
  \label{ineqbettertb1} \\ 
\frac{ \overline{m_b}}{ \overline{m_t}^6 } &\geq&
\frac{ Y_b^0} {(Y_t^0)^6 } \frac{ \sqrt{ E_2}}{E_1^9} \frac{1}{ \tan \beta \sin^5 \beta} 
 (2 \sqrt{2}G_F)^{5/2} 
  \label{ineqbettertb2} \\ 
\nonumber
\end{eqnarray}

The different origin and meaning of these inequalities as compared to 
(\ref{tanbound}) should be clear. 

\subsection{A numerical illustration}

Even though the general form of the exact solutions is not directly
exploitable analytically, truncated iterations provide very
good approximations which can be furthermore very well controlled using
the convergence criteria we derived. For instance truncating at the
first iteration, {\sl i.e.} approximating $E_{12}$ and $E_{21}$
in Eqs.((\ref{e12}, \ref{e21}) by the explicit forms

\begin{eqnarray}
&&E_{12}(t)\simeq \frac{ E_1(t)}{ (1 - b y_b^0 \int_0^t E_{2}(t') dt')^{a/b}} 
\label{approxe12} \\
&& \nonumber \\
&&E_{21}(t) \simeq \frac{ E_2(t)}{ (1 - b y_t^0 \int_0^t E_{1}(t') dt')^{a/b} } 
\label{approxe21}  \\
\nonumber
\end{eqnarray}

and plugging them back into Eqs.(\ref{ytsol}, \ref{ybsol}), one gets a simple
analytical solution. In table 1, a comparison is made for this solution
with the Runge-Kutta method, showing an excellent agreement of less than 
1\% accuracy for any small, moderate or large values of $\tan \beta$.

\begin{table}[htb]
\begin{tabular}{|c|c|c|c|c|c|c|}
\hline\hline 
$\tan \beta$ &$Y_b^0$ & $Y_t^0$ & $Y_b(t)$, truncated&  $Y_b(t)$, R.-K.& $Y_t(t)$, truncated&  $Y_t(t)$, R.-K.\\ \hline\hline
2&0.0387453 &1.13007 & 0.0145059&0.0145050 &0.775788 &  0.775974               \\ \hline
10&0.174138 &1.01581 & 0.0630978 &0.0631052 &0.54263 & 0.542743                \\ \hline
50&0.866544 &1.01097 & 0.435682&0.439526 &0.585453 &    0.590258             \\ \hline \hline

\end{tabular}
\caption{Numerical comparison between the exact one-loop solution (truncated to the first iteration) and the Runge-Kutta RG evolution. The evolution is over 10 orders of magnitude starting from the initial Yukawa coupling values shown in
the table}
\end{table}
Similar approximations can be
obtained for the top-bottom-$\tau$ case, at least if the initial Yukawa
couplings verify the sufficient convergence criterion of section {\bf 3.2}.
We will not dwell on further possible applications in the present paper.       

\section{Conclusion}

We have written down integrated forms for the running of the
Yukawa couplings in the case of two and three Yukawa fermions,
which are easily generalizable to any number of such fermions.
These forms are {\sl exact} solutions for the one-loop renormalization
group equations, valid for virtually any gauge theory with a Yukawa sector. 
The most important feature of such forms is that they allow for a 
rigorous determination of convergence criteria as well as exact
conditions for avoiding Landau-like poles of the Yukawa couplings.
In the case of the MSSM, such criteria lead to approximate analytical 
solutions in the top-bottom system, with very good numerical accuracy 
( \lsim 1\%) for any value of $\tan \beta$. Similar criteria were
obtained for the top-bottom-$\tau$ system, which lead 
 to controllable analytical approximations.
In this context we gave some preliminary applications for Landau pole
bounds in the top-bottom system, commented on some infra-red fixed points and 
lines, and gave optimal necessary constraints on the
Yukawa couplings both in the top-bottom and top-bottom-$\tau$ systems.

In view of the increasing
phenomenological interest for the large $\tan \beta$ scenario, such
solutions should prove useful in determining the exact structure
of the running of the remaining parameters of the MSSM using for instance
the method developed in \cite{kazakov1}, and possible implications
on the structure of the (stable) infra-red fixed points \cite{jack}.
Very recently, the authors of reference \cite{kazakov2} have addressed
similar issues, starting though from approximate solutions.\\ 


{\bf Acknowledgment:} We are indebted to J.-L. Kneur for providing us
with the numerical illustration presented in section {\bf 5.4} and thank him 
as well as C. Le Mou\"el for discussions. 
This work has been performed partly in the context of 
{\sl GDR-Supersym\'etrie} where preliminary results were published
in \cite{bezouhetal}.

\newpage

\renewcommand{\thesection}{Appendix}
\renewcommand{\theequation}{A.\arabic{equation}}
\renewcommand{\thesection}{A:}
\setcounter{equation}{0}
\section{Necessary and sufficient conditions for non singular evolutions}
\subsection{A necessary condition}
A necessary condition not to meet a singularity in running {\sl up}
from an initial energy scale $t=t_0$  to a given high energy scale $t=T<t_0$ is easy to establish. Indeed, if $y_{b,t}(t)$ are free 
from singularities in the interval $[T, t_0]$, then $E_{i j}(t; t_0)$ are 
necessarily positive for any $t$ in this interval, as can be seen from
eqs.(\ref{y0sol}) and the fact that $b<0$ and $y_{b,t}(t)>0$. It then follows 
from eqs.(\ref{e1221}) that

\begin{equation}
E_{12}(t, t_0) \geq E_1(t, t_0) \;\;\ \mbox{and} \;\;
E_{21}(t, t_0) \geq E_2(t, t_0)
\label{eqA1}
\end{equation}
since the denominators in eqs.(\ref{e1221}) are always smaller than one
(recall that $a/b > 0$ and $ t < t_0$). From the above considerations
one gets immediately the inequalities  
\begin{equation}
1 - |b| y_{t}(t_0) \int^{t_0}_t dt' E_{12}(t';t_0) \leq 
1 - |b| y_{t}(t_0) \int^{t_0}_t dt' E_{1}(t';t_0)
\label{eqA2}
\end{equation}

\begin{equation}
1 - |b| y_{b}(t_0) \int^{t_0}_t dt' E_{21}(t';t_0) \leq 
1 - |b| y_{b}(t_0) \int^{t_0}_t dt' E_{2}(t';t_0)
\label{eqA3}
\end{equation}
 for any $t$ in the interval $[T, t_0]$. 
Again, from 
eq.(\ref{y0sol}),
the lefthand side of eq.(\ref{eqA2}) should remain positive for any $t$ in 
the interval $[T, t_0]$, for $y_t(t)$ being free from singularities there. 
In particular $t=T$ gives the most significant condition, whence

\begin{equation}
1 - |b| y_{t}(t_0) \int^{t_0}_T dt' E_{1}(t';t_0) > 0 \label{eqA4}
\end{equation}
which is the necessary condition given in eq.(\ref{necess1}). 
One obtains similarly the second inequality in eq.(\ref{necess2}).

On the other hand, from (\ref{eqA3}) and the positivity of its left-hand side
one gets

\begin{equation}
E_{12}(t;t_0) = \frac{E_1(t;t_0)}{( 1 - b y_b(t_0) \int_{t_0}^{t} 
E_{21}(t';t_0) dt')^{a/b}} \geq \frac{E_1(t;t_0)}{( 1 - b y_b(t_0) \int_{t_0}^{t} 
E_{2}(t';t_0) dt')^{a/b}} \label{eqA5}
\end{equation}

valid since $E_1$ and $a/b$ are both positive. One can thus repeat the same
proof which lead to (\ref{eqA4}) with $E_1(t;t_0)$ replaced by

$$ \frac{E_1(t;t_0)}{( 1 - b y_b(t_0) \int_{t_0}^{t} 
E_{2}(t';t_0) dt')^{a/b}} $$
in eq.(\ref{eqA1}) to get the first inequality in eq.(\ref{necess21}), and the
second in a similar way. The infinite tower of inequalities (\ref{necess2}-
\ref{necess2inf}) is obtained recursively in the same way.

\newpage

\subsection{A sufficient condition}  
Similarly to what was done in section 3, eqs.(\ref{e1221}) define a mapping
 ${\cal A}$ in the form

\begin{eqnarray}
e_{12}'(t) = \frac{1}{( 1 - |b| y_b^0 \int_t^{t_0} e_2(t') e_{21}(t') dt')^{a/b}} \nonumber
\\
e_{21}'(t) = \frac{1}{( 1 - |b| y_t^0 \int_t^{t_0} e_1(t') e_{12}(t') dt')^{a/b}} \nonumber \\
\label{mapping1}
\end{eqnarray}

where

\begin{eqnarray}
e_{ij}(t) &\equiv& \frac{ E_{ij}(t;t_0)}{E_i(t;t_0)} \label{def1} \\
e_i(t) &\equiv& E_i(t;t_0) \label{def2} \\
y^0_{t, b} &\equiv& y_{t,b}(t_0)  \label{def3} 
\end{eqnarray}
Again, we collect the $e_{ij}(t)$'s in a vector
\begin{equation}
\vec{E}(t)= \left(
\begin{array}{l}
e_{12}(t) \\
e_{21}(t) 
\end{array} \right)
\label{Evect}
\end{equation}

and consider the range $T \leq t \leq t_0$ where $t_0$ corresponds to some
low energy  scale at which initial values for $y_t, y_b$ are chosen, and T  
a high energy scale (typically a GUT scale) up to which we require the
Yukawa couplings to have a non singular behaviour. We also define a norm
similar to eq.(\ref{normdef})

\begin{equation}
\parallel \vec{E} \parallel = \max \{ \sup_{T \leq t \leq t_0} \mid e_{12}(t) 
\mid,
\sup_{T \leq t\leq t_0} \mid e_{21}(t)\mid \}
\label{normdef1}
\end{equation}

To determine the conditions we are looking for to  avoid singularities 
in the range $[T, t_0]$  it will actually suffice to ask when does the mapping
defined in eq.(\ref{mapping1}) become a contraction. Parts of the the proof 
will resemble that of section 3.1.  However, in contrast to the latter
case where the mapping defined in eq.(\ref{mapping}) could not have
singularities as long as $U_1(t), U_2(t) \geq 0$, in the present case
one has to make sure that the mapping ${\cal A}$ keeps $e_{ij}(t)$ within
a finite interval $ 1\leq e_{ij}(t) \leq R$.

For a given $R$, let us thus denote by $X_R$ the set of all vectors 
$\vec{E}(t)$ such that $ 1\leq e_{ij}(t) \leq R$ for any t in the interval 
$[T, t_0]$. We look for conditions on the values of $R, y_t^0, y_b^0 $ such that

\begin{itemize}
\item{{\sl i)}} The mapping ${\cal A}$ sends any element of $X_R$ in $X_R$,
(so that the $e_{ij}(t)$'s  remain in the interval $[1, R]$ 
after an arbitrary
number of iterations of ${\cal A}$.)
\item{{\sl ii)}} ${\cal A}$ is a strictly contracting mapping in $X_R$, that is
\begin{equation}
\parallel {\cal A}(\vec{E_1}) - {\cal A}(\vec{E_2})\parallel \leq K_R 
\parallel \vec{E_1} - \vec{E_2}\parallel
\label{contraction1}
\end{equation}
with some $ K_R <1$.
\end{itemize}

Condition {\sl i)} means that

\begin{equation}
e_{ij}'(t) = \frac{1}{( 1 - |b| y_{b,t}^0 \int_t^{t_0} e_j(t') 
e_{ji}(t') dt')^{a/b}} \leq R \label{eqi1}
\end{equation}

On the other hand, one finds from $ e_{ij}(t') \leq R$ that 

\begin{equation}
e_{ij}' \leq \frac{1}{( 1 - |b| y_{b,t}^0  R \int_t^{t_0} e_j(t')  dt')^{a/b}} \label{eqi2}
\end{equation}

provided that $ 1 - |b| y_{b,t}^0  R \int_t^{t_0} e_j(t')  dt'$ is a positive
number\footnote{ This condition will, however, turn out to be already
contained in the sufficient condition we are looking for.}.\\
 
In view of eqs.(\ref{eqi1}, \ref{eqi2}) a sufficient condition to obtain 
{\sl i)}
is

\begin{eqnarray}
\frac{1}{ (1 - |b| y_t^0 R \int_t^{t_0} e_1(t') dt')^{a/b}} &\leq& R 
\nonumber \\
& \mbox{and} & \nonumber \\ 
\frac{1}{ (1 - |b| y_b^0 R \int_t^{t_0} e_2(t') dt')^{a/b}} &\leq& R \nonumber 
\end{eqnarray}

which translates easily into

\begin{eqnarray}
y_1 &\leq& \frac{1}{R} - \frac{1}{R^{ 1 + 1/c}} \nonumber \\
y_2 &\leq& \frac{1}{R} - \frac{1}{R^{ 1 + 1/c}} \nonumber \\
\label{condi}
\end{eqnarray}

where $c \equiv a/b$ and

\begin{equation}
y_{{}^1_2} \equiv |b| y^0_{{}^t_b} \int_T^{t_0} e_{{}^1_2 }(t) dt
\label{defA}
\end{equation}

At this level $R$ is still an arbitrary number. However, the optimal situation 
would be to chose
it such that the upper bound (\ref{condi}) be the largest possible. 
This would be the case for
$R=(1 + 1/c)^c$, but one still has to check whether this value is 
compatible with the second requirement {\sl ii)} which we turn to now.

Using the inequality (\ref{ineq1}), one gets from eq.(\ref{mapping1})

\begin{eqnarray}
\mid e'_{12}(t) - \tilde{e}'_{12}(t) \mid &\leq&
\frac{c |b| y_b^0 \int_T^{t_0} dt' e_2(t')  \mid e_{21}(t') - \tilde{e}_{21}(t') \mid}{
[ 1 - |b| y_b^0 \max \{\int_T^{t_0} dt' e_2(t') e_{21}(t'), 
\int_T^{t_0} dt' e_2(t') \tilde{e}_{21}(t')\} ]^{1+c} } \nonumber \\
&\leq & \frac{c |b| y_b^0  \int_T^{t_0} dt' e_2(t')}{
[1 - |b| y_b^0 R \int_T^{t_0}  e_2(t') dt']^{1+c}} \parallel \vec{E} - \tilde{\vec{E}}
\parallel \nonumber \\
\label{ineqA} 
\end{eqnarray}
 valid for any $t$ in the interval $[T, t_0]$ and $\vec{E}, \tilde{\vec{E}}$
belonging to $X_R$. A similar inequality holds obviously for 
$\mid e'_{21}(t) - \tilde{e}'_{21}(t) \mid$, 
and one finally gets

\begin{equation}
\parallel {\cal A}(\vec{E}) - {\cal A} (\tilde{\vec{E}}) \parallel
\leq K_R \parallel \vec{E} - \tilde{\vec{E}} \parallel
\label{contract2}
\end{equation}

with 
\begin{equation}
K_R= \max \{ \frac{c y_2}{ (1 - R y_2)^{1+c}}, \frac{c y_1}{ (1 - R y_1)^{1+c}} \}
\end{equation}

where $y_1, y_2$ are defined in eq.(\ref{defA}). It is now easy to check
that when condition (\ref{condi}) is satisfied with strict inequalities,
one gets 
\begin{equation}
K_R < 1 \label{strict}
\end{equation}

even for the value of $R$ quoted before,  $R_0= (1 + 1/c)^c$,
which maximizes the bounds in eq.(\ref{condi}).
Since eqs.(\ref{contract2}, \ref{strict}) mean that the mapping is indeed 
contracting, one concludes that the sufficient conditions for {\sl i)} given
in eq.(\ref{condi}) with maximal bounds, i.e.

\begin{eqnarray}
y_1 < \frac{1}{c ( 1 + 1/c)^{1 + c}} \nonumber \\
y_2 < \frac{1}{c ( 1 + 1/c)^{1 + c}}  \nonumber \\
\label{suffice}
\end{eqnarray}

imply also {\sl ii)}. It follows that when (\ref{suffice}) (equivalently
(\ref{suffice1}, \ref{suffice2}) )
are satisfied, a unique, regular, solution for the eqs.(\ref{e1221}) exists in 
$X_{R_0}$.
 The regularity of $y_t, y_b$ as given by
eq.(\ref{y0sol}) is then an immediate consequence.   

\newpage
\renewcommand{\theequation}{B.\arabic{equation}}
\renewcommand{\thesection}{B:}
\section{Approximate solutions for $y_t \gg y_b \neq 0$}
\setcounter{equation}{0}

To first order in $y^0_b$ one finds for $y_t$
 
\begin{equation}
y_t(t) = \frac{y_t^0 E_1(t)}{1 - b y_t^0 F_1(t)} \big [
1 + \frac{ a y_b^0}{1 - b y_t^0 F_1(t)} \int_0^t 
\frac{E_2(t')}{(1 - b y_t^0 F_1(t'))^{a/b-1}} dt' \big ]
\end{equation}

where 

\begin{equation}
F_1(t) \equiv \int_0^t E_1(t') dt'
\end{equation}

and where the solution for $y_b$ is given in eq.(\ref{approxyb}).

\renewcommand{\theequation}{C.\arabic{equation}}
\renewcommand{\thesection}{C:}
\section{Exact integrated forms for an arbitrary number of Yukawa couplings}
\setcounter{equation}{0}

Under the restriction of flavour-conserving Yukawa couplings, and assuming that
the Higgs fields sit in  representations such that  
the renormalization group equations
for the Yukawa couplings can be cast in the following form
at the one-loop level \cite{chengetal}:

\begin{equation}
\frac{d}{dt} y_i = y_i ( f_i(t) + \sum_j a_{i j}  y_j )
\end{equation}

where $i, j$ count the fermion fields, 
and $y_i$ denotes the square of the $i^{th}$ Yukawa coupling.

Then the exact solution for each $y_i$ reads:

\begin{equation}
y_i(t) = \frac{y_i^0 u_i}{1 - a_{ii} y_i^0 \int_0^t u_i}
\end{equation}

where the $u_i$'s are given by the implicit equations 

\begin{equation}
u_i(t)= \frac{ E_i(t)}{\defprod ( 1 - a_{jj} y^0_j \int_0^t u_j)^{a_{ij}/a_{jj}} }
\end{equation}

and $E_i(t) = e^{\int_0^t f_i(t') dt'}$.


\newpage
\begin{thebibliography}{99}
\bibitem{PR}
B. Pendleton and G.G. Ross, Phys.Lett. B98 (1981) 291
\bibitem{Hill}
C.T.Hill, Phys. Rev. D24 (1981), 691
\bibitem{EWB} 
L.E. Iba\~nez and G.G. Ross, Phys. Lett. B110 (1982) 215;
K.Inoue, A. Kakuto, H. Komatsu and S. Takeshita, Prog. Theor. Phys. 68 (1982) 
927; 71 ( 1984) 413;
L. Alvarez-Gaum\'e, M. Claudson and M.B. Wise, Nucl. Phys. B207 (1982) 96;
J. Ellis, D.V. Nanopoulos and K. Tamvakis, Phys. Lett B121 (1983) 123;
L.E. Iba\~nez, Nucl. Phys. B218 (1983) 514; 
J. Ellis, J.S. Hagelin, D.V. Nanopoulos and K. Tamvakis, Phys. Lett. B125 
(1983) 275; 
L.E. Iba\~nez and C. Lopez, Phys. Lett. B126 (1983) 54; Nucl. Phys. B236 (1984)
438
\bibitem{Alvarez}
L. Alvarez-Gaum\'e, J. Polchinski and M.B. Wise, Nucl. Phys. B221 (1983) 495
\bibitem{BHL}
W.A. Bardeen, C.T. Hill and M. Lindner, Phys. Rev. D41 (1990) 1647
\bibitem{Review} For reviews see: H.P.~Nilles, Phys. Rep. 110, 1 (1984); \\ 
H.E.~Haber and G.L.~Kane, Phys. Rep. 117, 75 (1985);
\bibitem{spectrum}
D.J. Casta\~no, E.J. Piard and P. Ramond, Phys.Rev.D49 (1994) 4882;
W. de Boer, R. Ehret and D.I. Kazakov, Z. Phys. C67 (1994) 647;
V. Barger, M.S. Berger and P. Ohmann, Phys.Rev.D49 (1994) 4908;
\bibitem{casas} J.A. Casas, A. Lleyda, C. Mu\~noz,
Nucl. Phys. B471 (1995) 3, and references therein
\bibitem{solutions}
L.E. Ib\'a\~{n}ez and C. Lop\'ez, Phys.Lett. 126B (1983) 54; 
Nucl. Phys. B233 (1984) 511; {\sl ibid.} B256 (1985) 218
\bibitem{Arnowitt}
P. Nath and  R. Arnowitt Phys.Rev. D56 (1997) 2820
\bibitem{btau}
M.S. Chanowitz, J. Ellis and M.K. Gaillard, Nucl. Phys. B128 ( 1977) 506;
A.J. Buras, J. Ellis, M.K. Gaillard and D.V. Nanopoulos,
Nucl. Phys. B135 (1978) 66; for more references see for instance 
\cite{schrempp};
\bibitem{EL1}
E.G. Floratos, G.K. Leontaris, Phys.Lett. B336 (1994) 194
\bibitem{EL2}
E.G. Floratos, G.K. Leontaris, Nucl.Phys. B452 (1995) 471;
E.G. Floratos, G.K. Leontaris and S. Lola, Phys.Lett. B365 (1996) 149
\bibitem{falck}
N.K. Falck, Z. Phys. C30 (1986) 247
\bibitem{diffeq}
see for instance ``Ordinary Differential Equations and Their Solutions'',
G.M. Murphy, Van Nostrand Reinhold Ed. (1960)
\bibitem{chengetal}
 T.P. Cheng, E. Eichten and L.-F. Li, Phys. Rev. D9 (1974) 2259
\bibitem{schrempp}
B. Schrempp and M, Wimmer, Prog. Part. Nucl. Phys. 37 (1996) 1
and references therein;
\bibitem{schrempp2}
B. Schrempp, Phys.Lett. 344 (1995) 193
\bibitem{kazakov1}
D.I. Kazakov, Phys. Lett. B449 (1999) 201
\bibitem{jack}
I. Jack, D.R.T. Jones Phys. Lett. B443 (1998) 177 
\bibitem{kazakov2}
S. Codoban and D.I. Kazakov, hep-ph/9906256
\bibitem{bezouhetal}
MSSM Working Group, PM-98-45, hep-ph/9901246 
\end{thebibliography}
\end{document}